\documentclass[reprint,
superscriptaddress,aps,showpacs,preprintnumbers,prd,notitlepage,notitlepage,amsmath,amssymb,prb,twocolumn,superscriptaddress,showpacs]{revtex4-1}
\usepackage{amsfonts}
\usepackage{graphicx, bm, amsmath, amsfonts, amssymb}
\usepackage{tabularx}
\usepackage{caption}
\usepackage{subcaption}
\graphicspath{{figures/}}
\usepackage{slashed}

\usepackage{comment}
\usepackage{graphicx}

\usepackage[usenames,dvipsnames]{xcolor}
\usepackage[colorlinks, breaklinks=true, linkcolor=red, citecolor=blue, linktocpage=true]{hyperref}
\usepackage{mathrsfs}
\usepackage{pst-node}
\usepackage{tikz-cd} 
\usepackage{cancel}
\usepackage{extarrows}

\newcommand{\LL}{\langle}
\newcommand{\RR}{\rangle}

\begin{document}
\title{The emergence of Einstein gravity from topological supergravity in $3+1$D}

\author{Tianyao Fang}
 \affiliation{Department of Physics, The Chinese University of Hong Kong, Shatin, New Territories, Hong Kong, China}

\author{Zheng-Cheng Gu}
\email{zcgu@phy.cuhk.edu.hk}
 \affiliation{Department of Physics, The Chinese University of Hong Kong, Shatin, New Territories, Hong Kong, China}

\begin{abstract}
The topological aspects of Einstein gravity suggest that topological invariance could be a more profound principle in understanding quantum gravity. In this work, we explore a topological supergravity action that initially describes a universe without Riemann curvature, which seems trivial.
However, we made a surprising discovery by introducing a small deformation parameter $\lambda$, which can be regarded as an AdS generalization of supersymmetry (SUSY). We find that the deformed topological quantum field theory (TQFT) becomes unstable at low energy, resulting in the emergence of a classical metric, whose dynamics are controlled by the Einstein equation. Our findings suggest that a quantum theory of gravity could be governed by a UV fixed point of a SUSY TQFT, and classical spacetime ceases to exist beyond the Planck scale.  
\end{abstract}

\maketitle
\textit{Introduction} --- Recent developments on topological aspects of the early universe at extremely high energy scales have opened up the possibility of describing quantum gravity using topological theory\cite{early1,early2,early3,early4}. This approach is characterized by being background independent and devoid of local degrees of freedom. One of the advantages of topological theory is its inherent ability to address puzzles related to homogeneity, isotropy, and scale invariance\cite{early1}. On the other hand, recent research suggests that the emergence of 
Einstein-Cartan action at low energy can also be attributed to the principle of topological invariants, which naturally excludes all higher-order terms, rather than the commonly known principle of general covariance. Moreover, a topological fixed point at UV scale also naturally resolves the long-standing renormalizability problem for quantum gravity. 


Historically, the topological nature of quantum gravity becomes manifest in the $2+1$D case, where the absence of local degrees of freedom plays a crucial role. Extensive efforts have been dedicated to the exact solution of $2+1$D quantum gravity, including the development of Virasoro topological quantum field theory (TQFT)\cite{witten3d, vtqft}.
In $3+1$D, topological gravity was initially proposed by Witten, describing a self-dual Weyl action\cite{early4}. However, due to the BRST symmetry, the Einstein action cannot be generated.
Inspired by Witten\cite{CS1}, Chamseddine developed a topological gravity theory in 2n+1 dimensions using the Chern-Simons form\cite{tg1,tg2,tg3,tg4}. 
In recent years, the replacement of the Chern-Simons form with transgression forms as a generalization has been extensively studied\cite{trans1,trans2,trans3,trans4,trans5,trans8,trans9,trans10}, promoting the quasi-invariant action to gauge invariant. 



In the past decade, there has been significant progress in the study of topological phases of quantum matter, leading to a systematic understanding of TQFT in $3+1$D. Specifically, it has been proposed that twisted BF theory is closely related to Einstein gravity. However, there remains a lack of controlled methods for deriving Einstein gravity from a TQFT fixed point.
In this paper, we aim to generalize the 3+1D topological gravity theory by incorporating SUSY. We begin by considering the simplest case of the $N=1$ topological supergravity theory in $3+1$D, without a cosmological constant term. We find that such a theory describes a trivial universe with zero Riemann curvature.
Surprisingly, by introducing a deformation parameter, denoted as $\lambda$, which can be viewed as an analogue of the Anti-de Sitter (AdS) generalization of SUSY transformations, we discover that such a super TQFT becomes unstable, leading to the emergence of classical spacetime and the Einstein-Cartan action with negative cosmological constant at low energy! Our results indicate that classical spacetime ceases to exist beyond the Planck scale.

 
\textit{Topological supergravity} --- As a warm up, we consider the following topological invariant SUSY action:
\begin{eqnarray}
S_{\text{Top}}&=& -\frac{1}{12} \int \varepsilon_{abcd}  e^{a}\wedge\overline{\psi} \gamma^{bcd}\wedge D\psi +\frac{1}{2 }\int \overline{c}\wedge D\psi \nonumber\\
&+& \frac{1}{2}\int \widetilde{B}_{a} \wedge  \left(T^{a}-j^{a}\right) +\frac{1}{2}\int B_{ab} \wedge R^{ab} . \label{snew1}
\end{eqnarray}
where $e^{a},\omega^{ab}$ and $\psi$ are 1-form fields, known as vierbein, spin connection, and gravitino. $j^a=\frac{1}{4}\overline{\psi} \gamma^{a}\wedge \psi$ is the gravitino current where the definition of Majorana conjugation $\overline{\psi}$ and $\overline{c}$ can be found in Supplementary Material. The torsion and (super) curvatures read:
\begin{eqnarray}
&&R^{ab}=d\omega^{ab}+\omega^{a}_{\ c}\wedge\omega^{cb},\ \ T^{a}\equiv De^{a}=de^{a}+\omega^{a}_{\ b}\wedge e^{b},   \nonumber\\
&& D\psi=d\psi+\frac{1}{4}\gamma^{ab}\omega_{ab}\wedge\psi.
\end{eqnarray}
The 2-form fields $\widetilde{B}^{a},B^{ab}$ and $c$ (which is a real Grassmann field) play the role of Lagrangian multiplier to compensate for the variation of the SUSY transformation from the first term in Eq. (\ref{snew}).  It is easy to check that the above action $S_{\text{Top}}$ is invariant under the following SUSY transformation: 
\begin{eqnarray}
&&\delta\psi=D\epsilon, \ \ \ \delta e^{a}=-\frac{1}{2}\overline{\psi}\gamma^{a}\epsilon,\ \ \  \delta\widetilde{B}_a=\frac{1}{6}\varepsilon_{abcd}\overline{\epsilon}\gamma^{bcd}D\psi,\nonumber\\
&&\delta B^{ab}=-\frac{1}{4}\overline{c}\gamma^{ab}\epsilon+\frac{1}{2}\varepsilon^{abcd}e_{c}\wedge \overline{\epsilon}\gamma_{d}\psi,\ \ \ \delta c=\frac{1}{2}\widetilde{B}_a\gamma^{a}\epsilon. \nonumber\\
 \label{SUSY1}
\end{eqnarray}
See Supplementary Material for more details. Variation with respect to the Lagrangian multiplier fields $\widetilde{B}^{a},B^{ab}$ and $c$ gives rise to the following: 
\begin{eqnarray}
    R^{ab}=0,\ \ \ T^{a}=j^{a},\ \ \ D\psi=0.
\end{eqnarray}
Similarly to the topological gravity theory without SUSY, such a theory also describes a trivial universe with vanishing classical metric and (super) curvatures. In the Supplementary Material, we also provide a full quantum treatment for the action Eq. (\ref{snew1}) to support this statement.

\textit{Deformed topological supergravity} --- Now we consider the most general topological supergravity theory which includes small regulator terms.
Such a SUSY TQFT can be regarded as a deformed theory of Eq. (\ref{snew1}) with deformation parameter $\lambda$.
\begin{widetext}
\begin{eqnarray}
S^\prime_{\text{Top}}&=& \int \varepsilon_{abcd} \left(  -\frac{1}{12} e^{a}\wedge\overline{\psi} \gamma^{bcd}\wedge D\psi +\frac{\lambda}{8}e^{a}\wedge e^{b}\wedge\overline{\psi} \gamma^{cd}\wedge \psi-\frac{\lambda^{2}}{2} e^{a}\wedge e^{b}\wedge e^{c}\wedge e^{d}\right) + \frac{1}{2}\int \widetilde{B}_{a} \wedge  \left(T^{a}-j^{a}\right)  \nonumber\\
&+&\frac{1}{2}\int B_{ab} \wedge \left( R^{ab}+\frac{\lambda}{2}\overline{\psi} \gamma^{ab}\wedge \psi+4\lambda^{2}e^{a}\wedge e^{b}\right) +\frac{1}{2 }\int \overline{c}\wedge \left( D\psi-\lambda \gamma^{a}\psi\wedge e_{a}\right).\label{snew}
\end{eqnarray}
 We note that $\lambda$ is a dimensionless parameter that plays the role of a regulator. The above action $S_{\text{Top}}$ is invariant under the following SUSY transformation:
\begin{eqnarray}
&&\delta\psi=D\epsilon+\lambda e^{a}\gamma_{a}\epsilon, \ \ \ \delta e^{a}=-\frac{1}{2}\overline{\psi}\gamma^{a}\epsilon,\ \ \  \delta\omega^{ab}=\lambda \overline{\psi}\gamma^{ab}\epsilon, \ \ \ \delta\widetilde{B}^a=-\Gamma^{a}(\epsilon)-\lambda\overline{c}\gamma^{a}\epsilon,\nonumber\\
&&\delta B^{ab}=-\frac{1}{4}\overline{c}\gamma^{ab}\epsilon+\frac{1}{2}\varepsilon^{abcd}e_{c}\wedge \overline{\epsilon}\gamma_{d}\psi,\ \ \ \delta c=\frac{1}{2}\widetilde{B}_a\gamma^{a}\epsilon-\lambda B_{ab}\gamma^{ab}\epsilon+\frac{\lambda}{4}\varepsilon_{abcd}e^{a}\wedge e^{b}\gamma^{cd}\epsilon, \label{SUSYtrans}
\end{eqnarray}
\end{widetext}
where $\Gamma_{a}(\epsilon)=\frac{1}{6}\varepsilon_{abcd}(-\overline{\epsilon}\gamma^{bcd}D\psi+\lambda\overline{\epsilon}\gamma^{bcdf}\psi\wedge e_{f}-3\lambda e^{b}\wedge\overline{\epsilon}\gamma^{cd}\psi)$.
See Supplementary Material for more details. In addition to local SUSY, the above action also possesses the following 1-form gauge transformation:
\begin{eqnarray}
\delta_{\widetilde{B}} \widetilde{B}^a&=&D\xi^{a},\ \ \ \delta_{\widetilde{B}} B^{ab}=-\frac{1}{2}(\xi^{a}\wedge e^{b}-\xi^{b}\wedge e^{a}),\nonumber\\
  \delta_{\widetilde{B}} c &=&\frac{1}{2}\xi_a\gamma^{a}\wedge\psi, \label{gaugetrans3}\\
\delta_{B} \widetilde{B}^a&=&-8\lambda^{2}\varsigma^{ab}\wedge e_{b},\ \ \ \delta_{B} B^{ab}=D\varsigma^{ab},\nonumber\\
\delta_{B} c&=&-\lambda\varsigma_{ab}\gamma^{ab}\wedge\psi,\label{gaugetrans4}\\
\delta_{c}\widetilde{B}&=&-\lambda\overline{\tau}\gamma^{a}\wedge\psi,\ \ \ \delta_{c} B^{ab}=-\frac{1}{4}\overline{\tau}\gamma^{ab}\wedge\psi,\nonumber\\
\delta_{c} c&=&D\tau+\lambda\gamma_{a}e^{a}\wedge\tau,\label{gaugetrans5}
\end{eqnarray}
where $\xi^{a},\varsigma^{ab}$ are 1-form bosonic gauge parameter and $\tau$ is 1-form Grassmann spinor. 

\textit{Second order formalism} -- 
We first incorporate over $\widetilde{B}$:
\begin{equation}
\int_{-\infty }^{+\infty}[D\widetilde{B}]\exp[\frac{i}{2 }\int \widetilde{B}_{a} \wedge  \left(T^{a}-j^{a}\right)],
\end{equation}
which leads to the delta function 
$\delta(T^{a}-j^{a})$.
Very different from the second-order formalism of the usual $N=1$ supergraivty theory, here the condition $T^{a}=j^{a}$ is imposed as a quantum constraint, rather than the classical equation of motion.
However, similarly to the usual case, the solution of $\omega$ for the above constraint can be obtained by splitting $\omega$ into:
\begin{equation}
\omega^{ab}=\Gamma^{ab}+K^{ab}, \label{Christofell}
\end{equation}
where $\Gamma^{ab}$ is the torsion free Christofell connection which depends only on $e^{a}_{\mu}$ and satisfies:
\begin{equation}
de^{a}+\Gamma^{a}_{\ b}\wedge e^{b}=0.
\end{equation}
Thus, $\Gamma^{a}_{\ b\mu}$ can be obtained as usual:
\begin{eqnarray}
    \Gamma_{\ b \mu}^{a}&=&\frac{1}{2}[e^{\rho}_{b}(\partial _{\rho }e_{\mu}^{a}-\partial _{\mu }e_{\rho}^{a})+e^{a\rho}(\partial _{\mu }e_{b\rho}-\partial _{\rho}e_{b\mu})\nonumber\\
&&+e^{a\lambda}e^{\nu}_{b}e^{c}_{\mu}(\partial _{\nu }e_{c\lambda}-\partial _{\lambda }e_{c\nu})],
\end{eqnarray}
where $e_b^\rho$ is the inversion of $e_\mu^b$ satisfying $e_b^\mu e_\mu^a=\delta^{a}_b, e_a^\rho e_\mu^a=\delta^{\rho}_\mu$. At this stage, we do not worry about the irreversibility of $e$. We will limit the discussion to the case where $e$ can be expanded perturbatively in a classical background throughout the whole paper. Moreover, the 1-form contorsion $K^{ab}$ satisfies:
\begin{equation}
T^{a}=K^{a}_{\ b}\wedge e^{b}=j^{a}, 
\end{equation}
which can be solved as:
\begin{eqnarray}
K_{ab\mu}=-\frac{1}{4}\left(e^{\rho}_{b}\overline{\psi}_{\rho} \gamma_{a}\psi_{\mu}-e^{\rho}_{a}e^{\sigma}_{b}e^{c}_{\mu}\overline{\psi}_{\rho} \gamma_{c}\psi_{\sigma}+e^{\rho}_{a}\overline{\psi}_{\mu} \gamma_{b}\psi_{\rho}\right).\nonumber\\ \label{contorsion}
\end{eqnarray}
Using the decomposition Eq. (\ref{Christofell}), we can rewrite $R_{ab}$ as:
\begin{equation}
R_{ab}=\widetilde{R}_{ab}+\widetilde{D}K_{ab}+K_{ac}\wedge K^{c}_{\ b},
\end{equation}
where $\widetilde{D}$ is the covariant derivative with respect to $\Gamma^{ab}$ and $\widetilde{R}^{ab}=d\Gamma^{ab}+\Gamma^{a}_{\ c}\Gamma^{cb}$ is the torsion free Riemann tensor.

After eliminate $\widetilde{B}^a$ and $\omega^{ab}$, $S^\prime_{\text{Top}}$ becomes:
\begin{eqnarray}
S&=&  -\frac{1}{2}\int d^{4}x\ e ( \overline{\psi}_{\mu} \gamma^{\mu\nu\rho}D_{\nu}\psi_{\rho} -\lambda \overline{\psi}_{\mu} \gamma^{\mu\rho} \psi_{\rho}+24\lambda^{2})\nonumber\\
&+&\frac{1}{8}\int d^{4}x\  \varepsilon^{\mu\nu\rho\sigma} B_{ab\mu\nu}( R^{ab}_{\rho\sigma}+\lambda\overline{\psi}_{\rho} \gamma^{ab}\psi_{\sigma}+8\lambda^{2}e^{a}_{\rho} e^{b}_{\sigma})\nonumber\\
&+&\frac{1}{4 }\int d^{4}x\  \varepsilon^{\mu\nu\rho\sigma}\overline{c}_{\mu\nu} ( D_{\rho}\psi_{\sigma}-\lambda \gamma^{a}\psi_{\rho} e_{a\sigma}),\label{safter}
\end{eqnarray}
where $e=\det e^{a}_{\mu}$ and $\gamma^{\mu}=e^{\mu}_{a}\gamma^{a}$.  $\omega$ is expressed with respect to $e$ and $\psi$. We used identities:
\begin{eqnarray}
&&\varepsilon_{abcd}e^{a}_{\mu}=e\varepsilon_{\mu\nu\rho\sigma}e^{\nu}_{b}e^{\rho}_{c}e^{\sigma}_{d},\nonumber\\
&& \varepsilon_{\mu\nu\rho\sigma}\varepsilon^{\mu'\nu'\rho'\sigma}=\delta^{\mu'}_{\mu}\delta^{\nu'}_{\nu}\delta^{\rho'}_{\rho}\pm \text{permutations of $\mu,\nu,\rho$}.\nonumber
\end{eqnarray}
We also use the convention $\varepsilon_{0123}=1$ for Lorentz index $a$ and $\varepsilon^{0123}=1$ for spacetime index $\mu$.
Now $S$ becomes a second order formalism, which is still SUSY invariant with suitable modification for the variation of $c_{\mu\nu}$:
\begin{eqnarray}
    \delta \overline{c}_{\mu\nu}&=&-\frac{1}{4}\widetilde{B}_{a\mu\nu}\overline{\epsilon}\gamma^{a}+\frac{1}{2}\lambda B_{ab\mu\nu}\overline{\epsilon}\gamma^{ab}\nonumber\\
&& +\frac{1}{4}\varepsilon_{\mu\nu\rho\sigma}(F^{ab\sigma}\overline{\epsilon}\gamma_{a}e^{\rho}_{b}-\frac{1}{2}F^{ab\lambda}\overline{\epsilon}\gamma_{\lambda}e^{\rho}_{a}e^{\sigma}_{b}),
\end{eqnarray}
where $F^{ab\sigma}=\frac{1}{4}\varepsilon^{\mu\nu\rho\sigma}(\overline{c}_{\mu\nu}\gamma^{ab}\psi_{\rho}+D_{\rho}B^{ab}_{\ \ \mu\nu})$. SUSY transformations of $e^a,\psi$ and $B^{ab}$ remain unchanged. The variation of $\omega^{ab}$ is obtained using the chain rule. In addition, the 1-form gauge transformations of $B,c$ in Eq. (\ref{gaugetrans4}) and  Eq. (\ref{gaugetrans5}) remain unchanged. We drop the subscript "Top" since topological invariance is not manifest now.

\textit{Low energy effective theory and saddle point approximation} --- We conjecture that the UV fixed point of quantum gravity is actually controlled by such a non-unitary TQFT which is unstable. 
In the low-energy limit, it will flow to the phase described by Einstein gravity with a nonzero vaucuum expectation value (VEV) of $e_\mu^a$.
\begin{equation}
\langle e^{a}_{\mu}\rangle=\frac{1}{l_{p}}\overline{e}^{a}_{\mu},\ \ \ e^{a}_{\mu}=\frac{1}{l_{p}}(\overline{e}^{a}_{\mu}+h^{a}_{\mu}).
\end{equation}
where $l_p$ plays the roll as a dimension 1 order parameter: in the high energy TQFT phase, $l_p$ goes to infinity and the classical spacetime does not exist; while in the low energy phase, $l_p$ becomes finite and classical spacetime will emerge. 
In general, $\overline{e}^{a}_{\mu}$ is a function that depends on the spacetime coordinate determined by self-consistent equations. 
$h^{a}_{\mu}$ is quantum fluctuation around the classical background with $\langle h^{a}_{\mu}\rangle=0$.


We define the dimensionless vierbein field as:
$
\widetilde{e}^{a}_{\mu}=l_{p}e^{a}_{\mu}=\overline{e}^{a}_{\mu}+h^{a}_{\mu},
$
which is related to the \textit{emergent metric} (not the background metric of the underlying manifold where the path integral is defined) via $\widetilde{e}^{a}_{\mu}\widetilde{e}^{b}_{\nu}\eta_{ab}=g_{\mu\nu}$.  For convenience, we just rename $\widetilde{e}^{a}_{\mu}$ to $e^{a}_{\mu}$ (i.e., $e^{a}_{\mu}$ is now a dimensionless field) without causing confusion. 
In this paper, we will consider emergent metric within  maximally symmetric spacetime, i.e.
\begin{eqnarray}
    ds^2=-(1+r^2/a^2)dt^2+(1+r^2/a^2)^{-1}dr^2+r^2d\Omega^2 ,\label{maximum_symmetric_space} \nonumber\\
\end{eqnarray}
where the parameter $a$ represents the radius of the spacetime. The cases $a^2<0$, $a^2>0$, and $a^2=\infty$ correspond to de Sitter spacetime, Anti-de Sitter spacetime, and flat spacetime, respectively. The Ricci tensor can be expressed as
$
    \overline{R}_{\mu\nu}=\Lambda \overline{g}_{\mu\nu}.
$
in a maximally symmetric spacetime, where the cosmological constant can be expressed in terms of the radius of the spacetime as $ \Lambda=-3/a^2$. Note that $\overline{R}_{\mu\nu}$ is the emergent background Ricci tensor w.r.t. background metric $\overline{g}_{\mu\nu}=\overline{e}^{a}_{\mu}\overline{e}_{a\nu}$.


In addition, we further assume that $B^{ab}$ can also acquire a non-zero VEV:
\begin{equation}
\langle B^{ab}_{\ \ \mu\nu}\rangle =\frac{B}{l_p^2}\varepsilon^{abcd}\overline{e}_{c\mu}\overline{e}_{d\nu}\label{Bvev}
\end{equation}
where $B$ is a constant solved from self-consistent equation. Thus, the leading order terms of $S_{B}$ (Here we neglect the fluctuations of $B_{ab}$ and $e^{a}$ for saddle point calculation) becomes:
\begin{eqnarray}
S_{B}&=&\frac{1}{8}\int d^{4}x\ \varepsilon^{\mu\nu\rho\sigma}B^{ab}_{\mu\nu}( R_{ab\rho\sigma}+\lambda\overline{\psi}_{\rho}\gamma_{ab}\psi_{\sigma}+8\frac{\lambda^{2}}{l_{p}^{2}}e_{a\rho}e_{b\sigma})  \nonumber\\
&\approx&\frac{B}{2l_p^2}\int d^{4}x\ \overline{e}(\overline{R}+\lambda\overline{\psi}_{\mu}\gamma^{\mu\nu}\psi_{\nu}+48\frac{\lambda^{2}}{l_{p}^{2}}), \label{S_B}
\end{eqnarray}
Here the gamma matrix with respect to spacetime index is defined as $\gamma_{\mu}=e^{a}_{\mu}\gamma_{a}$. 

To simplify the discussion and acquire one-loop effective action, we also neglect all gravitino interaction terms, which allows us to replace all covariant derivative $D$ acting on $\psi_{\mu}$ with torsion free total covariant derivative $\nabla$ defined as:
\begin{equation}
\nabla_{\mu}\psi_{\nu}=\partial_{\mu}\psi_{\nu}+\frac{1}{4}\gamma^{ab}\Gamma_{ab\mu}\psi_{\nu}-\Gamma^{\rho}_{\ \mu\nu}\psi_{\rho}.
\end{equation}
The spin connection $\Gamma^{\rho}_{\ \mu\nu}$ is solved from:
\begin{equation}
\nabla_{\mu}e^{a}_{\nu}=\widetilde{D}_{\mu}e^{a}_{\nu}-\Gamma^{\rho}_{\ \mu\nu}e^{a}_{\rho}=\partial_{\mu}e^{a}_{\nu}+\Gamma^{a}_{\  b\mu}e^{b}_{\nu}-\Gamma^{\rho}_{\ \mu\nu}e^{a}_{\rho}=0.
\end{equation}
We leave the discussion for the effect of higher order gravitino interaction terms in our future work.

In order to solve for the values of these order parameters, we neglect their fluctuations, integrate out the fermionic degrees of freedom to obtain the one-loop effective action, and then employ the self-consistent equations for the solution. With all these assumptions and simplifications, the original action Eq. (\ref{safter}) can be rewritten as:
\begin{eqnarray}
S'&=&\frac{B}{2l_p^2}\int \overline{e} \ (\overline{R}+\frac{24(2B-1)}{B}\frac{\lambda^2}{l_p^2})+S_f,  \nonumber\\
S_f&=& -\frac{1}{2l_p}\int \overline{e} \ \overline{\psi}_{\mu}(\gamma^{\mu\nu\rho}\nabla_{\nu}-m_{\psi}\gamma^{\mu\rho})\psi_{\rho}\nonumber\\
&& +\frac{1}{4}\int\varepsilon^{\mu\nu\rho\sigma}\overline{c}_{\mu\nu}(\nabla_{\rho}+m_{c}\gamma_{\rho})\psi_{\sigma}, \label{tree_and_oneloop}
\end{eqnarray}
where masses are defined as:
\begin{eqnarray}
    m_c=\lambda/l_p,\ \ \ m_\psi=(B+1)m_c.
\end{eqnarray}
The specific form of the one-loop effective action depends on the spacetime, i.e., the behavior of $a^2$. By solving the self-consistent equations, we find that flat spacetime and de Sitter space do not possess saddle points, details can be found in Supplementary Material.
Therefore, we only present the case for Anti-de Sitter spacetime in the following. The one-loop effective action is obtained by a Gaussian integration over $\psi_{\mu}$ and $c_{\mu\nu}$ as:
\begin{eqnarray}
    S_{\text{eff}}&=&-\frac{6V(H_4)}{l_p^2}[\frac{2(2B-1)\lambda^2}{l_p^2}-\frac{ B}{a^2}]   \nonumber\\ 
    &&-\frac{1}{2}\ln\frac{\det\triangle_{3/2}(-\frac{13}{6}\Lambda)}{\det\triangle_{3/2}(m_\psi^2)[\det\triangle_{3/2}(0)]^{3/2}}, \label{effect_action}
\end{eqnarray}
where $V(H_4)=\int d^4x \overline{e}$ is the volume of the hyperbolic space of 4 dimensions, we also used the relation between the radius $a$ of the Ads space of 4 dimensions and its cosmological constant of correspondent $\Lambda=-3/a^2$.  The definition of constraint operators and
more details of integration can be found in the Supplementary Material. The one-loop functional determinant can be determined by regularizing zeta functions\cite{ADSspecturm}:
\begin{eqnarray}
    \ln\det\frac{\triangle_{s}(X)}{\mu^2}=\zeta^{(s)}(0,a^2X)\ln(\frac{1}{|a^2|\mu^2})-\zeta^{(s)'}(0,a^2X), \nonumber
\end{eqnarray}
where $\mu$ is dimension 1 normalization parameter. We leave the explicit form of zeta functions in Supplementary Material. The values of $a$ and $B$ are then solved from self-consistent equations:
\begin{eqnarray}
\frac{\delta S_{\text{eff}}}{\delta l_p}=\frac{\delta S_{\text{eff}}}{\delta B}=\frac{\delta S_{\text{eff}}}{\delta \overline{e}^{a}_\mu}=0. \label{self-consistent}
\end{eqnarray}
Since the effective action depends on the $\overline{e}^{a}_\mu$ only through the background metric (volume element and Ricci tensor), its variation with respect to $\overline{e}^{a}_\mu$ is equivalent to the variation with respect to the metric $\overline{g}_{\mu\nu}$. We consider the variation only within the AdS spacetime, i.e. the variation of the metric only leads to a change in the radius of Ads spacetime. By using the relation between the Ricci tensor and the radius $R_{\mu\nu}=-3g_{\mu\nu}/a^2 $, the variation of radius $a$ with respect to metric can be obtained as:
\begin{eqnarray}
    \frac{\delta a}{\delta \overline{g}_{\mu\nu}}=\frac{\sqrt{12}}{(-\overline{R})^{3/2}}\frac{1}{2}\frac{\delta \overline{R}}{\delta \overline{g}_{\mu\nu}}= \frac{a^3}{24}\frac{\delta\overline{ R}}{\delta \overline{g}_{\mu\nu}}.
\end{eqnarray}
This form can be further simplified as: 
\begin{eqnarray}
     \frac{\delta a}{\delta \overline{g}_{\mu\nu}}=-\frac{a^3}{24}\overline{R}^{\mu\nu}+t.d.=\frac{a}{8}\overline{g}^{\mu\nu}+t.d.,
\end{eqnarray}
We note that $\delta \overline{R}=\delta \overline{g}^{\mu\nu} \overline{R}_{\mu\nu}+\overline{g}^{\mu\nu}\delta \overline{R}_{\mu\nu}$ where $\delta R_{\mu\nu}$ is a total derivative (denoted as $t.d.$) since $\zeta$ and $\zeta'$ are functions of $a$ only, and it does not contribute to self-consistent equations.
We leave the detailed process of solving the self-consistent equations to Supplementary Material,
and we only find one Ads solution for Eq. (\ref{self-consistent}):
\begin{eqnarray}
  B =7.52,\ \  \Lambda=-\frac{36\lambda^2}{l_p^2},\ \ \ \   \frac{1}{l_p^2}=  \frac{432\mu^2}{\lambda^2}  e^{- \frac{9.04}{\lambda^{2}} } .\label{non_trivial_solution}
\end{eqnarray}
 This solutions possess a Ricci curvature $R\sim \lambda^2/l_p^2$. The explicit form of $\overline{e}^a_\mu$ can be obtained from the Ads metric Eq. (\ref{maximum_symmetric_space}). We can choose a gauge such that the form of $\overline{e}^a_\mu$ is diagonalized.

\textit{The emergence of Einstein gravity} --- To this end, we see that although we start from a topological theory, a saddle point may emerge at low energy and the 2-form gauge field $B^{ab}$ may acquire a non-zero VEV with $\langle B^{ab}_{\ \ \mu\nu}\rangle =Bl_p^{-2}\varepsilon^{abcd}\overline{e}_{c\mu}\overline{e}_{d\nu}$. In the following we will investigate the quantum fluctuation around such a saddle point. In general,
$B^{ab}_{\ \ \mu\nu}$ can be expanded as:
\begin{eqnarray}
l_p^{2}B^{ab}_{\ \ \mu\nu}&=&B\varepsilon^{abcd}e_{c\mu}e_{d\nu}+\beta^{ab}_{\ \ \mu\nu}\nonumber\\
&=&B\varepsilon^{abcd}\overline{e}_{c\mu}\overline{e}_{d\nu}+2B\varepsilon^{abcd}\overline{e}_{c\mu}h_{d\nu} \nonumber\\
    &&+B\varepsilon^{abcd}h_{c\mu}h_{d\nu}+O(h^3)+O(\beta),
\end{eqnarray}
where $\beta^{ab}_{\ \ \mu\nu}$ is quantum fluctuation around the classical background. Strictly speaking, $\beta^{ab}_{\ \ \mu\nu}$ should include fluctuations of $e^{a}_{\mu}$. However, in order for both parts to manifest as general coordinate transformation covariant, we extracted the part containing $h$ in $\beta$ to recover $\overline{e}$ to $e$. Thus, $\beta$ is obviously a tensor under a general coordinate transformation. 
Here we can omit the fluctuation $\beta$ because it acquires mass$\sim \lambda/l_p$ and the average $\langle \beta^2\rangle^{1/2}/B $ will be suppressed by volume. A similar discussion can be found in Ref. \cite{stringtheory} by Polyakov. Therefore, the original flat-curvature constraint term $S_B$ Eq. (\ref{S_B}) now contributes an Einstein action term, which is:
\begin{eqnarray}
    \frac{B}{2l_p^2}\int e(R+\frac{48\lambda^2}{l_p^2}), \label{S_B_final}
\end{eqnarray}
where $R$ is the curvature with respect to the emerging metric $g_{\mu\nu}=e^a_\mu e_{a\nu} $.  The determinants brought about by $\psi,c$ and ghosts can be regularized using the heat kernel method\cite{heat1,heat2}, which contributes some $\overline{R}^2$ terms (including combinations of the Ricci tensor and the Riemann tensor).  At the same time, integrations of $\psi$ and $c$ induce quadratic and interaction terms of $h$. Since the general covariance can not be broken at any energy scale, the low-energy effective action should possess diffeomorphism invariance.  All possible local terms should be contributed by the higher-order expansions of $h$ from the Hilbert-Einstein action $R$ and its higher-order products. In other words, we can always replace $\overline{R}$ (obtained from the heat kernel method) in effective action with $R$. By power counting, these higher-order terms would be suppressed by $l_p^2$, thus the possible effective action reads:
\begin{equation}
    S_{eff}=\frac{1}{2l_p^2}\int e ( R-2\Lambda' +O(l_{p}^{2} R^2)),
\end{equation}
where we perform a constant conformal transformation $g\rightarrow B^{-1}g$ to normalize the overall factor and $\Lambda'=-36 \lambda^2 (Bl_p)^{-2}$. The above effective action is exactly the Einstein action with negative cosmological constant, which arises from an underlying SUSY TQFT with generalized symmetry at UV scale. 

\textit{Conclusion and discussion} --- In this paper, we propose a topological supergravity theory that may be considered as a natural candidate for the early universe and quantum gravity. We speculate that at an extremely high energy scale beyond the Planck energy, classical spacetime will vanish, and the vierbein field $e_\mu^a$ will have a zero VEV. Unlike the usual unitary TQFTs with vanishing beta functions, the topological supergravity theory we propose should be regarded as a non-unitary TQFT, which could be unstable in $3+1$D. A self-consistent saddle point calculation suggests that it will flow to the Hilbert-Einstein action at low energy with non-zero VEV for the $e$ and $B$ fields. 
Moreover, our scenario indicates that SUSY might already be broken at Planck energy scale.

The advantage of such a topological supergravity theory is that it equips with both topological invariance and local SUSY, and the action is uniquely defined (at least for the $N=1$ case) up to certain field redefinitions. Although we only demonstrate the simplest topological supergravity with $N=1$, we believe that the saddle-point approximation is still valid, and we will carefully study the large-$N$ cases in the future. On the other hand, our result is also consistent with Ads/CFT correspondence, since CFT might naturally arise on the boundary of bulk TQFT. Moreover, our work even naturally explains the emergence of Ads background and fundamental constant such as Planck length $l_p$.

\textit{Acknowledgement} --- We would like to thank Yongshi Wu, Xiao-Gang Wen and Hong Liu for helpful discussions. This work is supported by a grant from the Research Grants Council of the Hong Kong Special Administrative Region, China (Project No. AoE/P-404/18).
 
\appendix

\section{Clifford algebra}\label{clifford}

In order to define spinors, we need to utilize Clifford algebra. Clifford algebra (in 3+1 dimension) is described a set of $\gamma$-matrices satisfy the anti-commutation relations:
\begin{eqnarray}
\gamma_a \gamma_b+\gamma_b \gamma_a=2\eta_{ab}, \label{gamma}
\end{eqnarray}
where $\eta_{ab}$ is the metric of Minkowski spacetime taken as $\text{diag}(-1,1,1,1)$. The full Clifford algebra consists of the identity $\mathbf{1}$ and 4 generating elements $\gamma_{a}$, plus all independent matrices formed from products of the generators. Since symmetric products reduce to a product containing fewer $\gamma$-matrices by Eq. (\ref{gamma}), the new elements must be antisymmetric products. We thus define:
\begin{equation}
\gamma_{a_{1}...a_{r}}=\gamma_{[a_{1}}...\gamma_{a_{r}]},\ \ \ \ \text{e.g.}\ \ \ \ \gamma_{ab}=\frac{1}{2}(\gamma_{a}\gamma_{b}-\gamma_{b}\gamma_{a}).
\end{equation}
And the complete set of Clifford algebra can be denoted as:
\begin{eqnarray}
\{\Gamma^{A}=\mathbf{1},\gamma^{a},\gamma^{a_{1}a_{2}},\gamma^{a_{1}a_{2}a_{3}},\gamma^{a_{1}a_{2}a_{3}a_{4}}\},\nonumber\\
\{\Gamma_{A}=\mathbf{1},\gamma_{a},\gamma_{a_{2}a_{1}},\gamma_{a_{3}a_{2}a_{1}},\gamma_{a_{4}a_{3}a_{2}a_{1}}\}, \label{set}
\end{eqnarray}
index values satisfy the conditions $a_{1}<a_{2}<...<a_{r}$, lower and up by $\eta_{ab}$. There are $C_{r}^{4}$ distinct index choices at each rank $r$ (rank $r$ we mean the product of $r$ $\gamma$-matrices. For convenience, we denote them by $\Gamma^{(r)}$) and a total of 16 matrices. For convenience, we define the highest rank Clifford algebra element as:
\begin{equation}
\gamma_{5}=i\gamma_{0}\gamma_{1}\gamma_{2}\gamma_{3}.
\end{equation}
It has the following properties:
\begin{equation}
\gamma_{5}^2=\mathbf{1},\ \ \ \ \{\gamma_{5},\gamma_{a}\}=0,\ \ \ \ \varepsilon_{abcd}\gamma^d=i\gamma_{5}\gamma_{abc},
\end{equation}
where we take the convention $\varepsilon_{0123}=1$, the last one properties can be proved by considering the explicit component. There exists an unitary charge conjugate matrix $C$ satisfies:
\begin{equation}
(C\Gamma^{(r)})^{T}=-t_{r}C\Gamma^{(r)},\ \ \ \ t_{r}=\pm 1, \label{pm}
\end{equation}
 For 3+1D supergravity, we take the convention:
\begin{equation}
t_0=t_3=t_4=1,\ \ \ t_1=t_2=-1.
\end{equation}
The Majorana conjugate is defined as:
\begin{equation}
\overline{\lambda}=\lambda^T C,
\end{equation}
where $\lambda$ is arbitrary Grassmann 4-components spinor. The bilinears of two Majorana fields $\chi$ and $\lambda$ has below symmetriy property (Majorana flip):
\begin{equation}
\overline{\lambda}\gamma_{\mu_1...\mu_r}\chi=t_{r}\overline{\chi}\gamma_{\mu_1...\mu_r}\lambda.
\end{equation}	
For 1-form gravitino $\psi=\psi_{\mu}dx^{\mu}$, this implies: 
\begin{equation}
\overline{\psi} \wedge\psi=\overline{\psi} \gamma^{abc}\wedge\psi=\overline{\psi} \gamma^{abcd}\wedge\psi=0
\end{equation}
because exchange the position of 1-form field gives an extra minus sign. The non-vanishing gravitino bilinear are: $\overline{\psi}\gamma^{a}\wedge\psi$ and $\overline{\psi}\gamma^{ab}\wedge\psi$. The Majorana fermion satisfy the reality condition\cite{supergravity}:
\begin{equation}
\psi^{*}=B\psi,
\end{equation} 
where $B$ is the complex conjugate matrix $B=it_{0}C\gamma^{0}$.

The explicit form of $B$ and $C$ varies under different representation of gamma matrix. The two most commonly used representations are\cite{supergravity}  \textbf{Weyl representation} where
\begin{equation}
\gamma ^{0}=\left( 
\begin{array}{cc}
0 & 1 \\ 
-1 & 0 \end{array}
\right) , \ \ \gamma ^{i}=\left( 
\begin{array}{cc}
0 & \sigma^{i} \\ 
\sigma^{i} & 0 \end{array}
\right) ,
\end{equation}
with $B=\gamma^{0}\gamma^{1}\gamma^{3},C=i\gamma^{3}\gamma^{1}$ and \textbf{real representation} where
\begin{equation}
\gamma^{0}=i\sigma_{2}\otimes\mathbf{1},\ \ \gamma^{1}=\sigma_{3}\otimes\mathbf{1},\ \ \gamma^{2}=\sigma_{1}\otimes\sigma_{1},\ \ \gamma^{3}=\sigma_{1}\otimes\sigma_{3},
\end{equation}
with $B=\mathbf{1}$ up to a phase and $C=i\gamma^{0}$ respectively.

In the end of this part, we list two useful identities of gamma matrix:
\begin{eqnarray}
&&\gamma^{a_{1}...a_{n}}\gamma^{b}=\gamma^{a_{1}...a_{n}b}+\sum_{i=1}^{n}(-1)^{n+i}\gamma^{a_{1}...a_{i-1}a_{i+1}...a_{n}}\eta^{a_{i}b},\nonumber\\
&&\gamma^{b}\gamma^{a_{1}...a_{n}}=\gamma^{ba_{1}...a_{n}}+\sum_{i=1}^{n}(-1)^{1+i}\gamma^{a_{1}...a_{i-1}a_{i+1}...a_{n}}\eta^{a_{i}b},\nonumber\\
&&\gamma^{abc}\gamma_{mn}=-6\gamma^{[ab}_{\ \ [m}\delta^{c]}_{n]}-6\gamma^{[a}\delta^{b}_{[m}\delta^{c]}_{n]},\nonumber\\
&&\gamma_{mn}\gamma^{abc}=6\gamma^{[ab}_{\ \ [m}\delta^{c]}_{n]}-6\gamma^{[a}\delta^{b}_{[m}\delta^{c]}_{n]},\label{grassmannidentity2}
\end{eqnarray}
where the bracket $[,]$ denotes the antisymmetrization of the indices. The first one can be proved by rewritting the gamma matrix as:
\begin{equation}
\gamma^{a_{1}...a_{n}}=\gamma^{a_{1}}...\gamma^{a_{n}},\ \ \ \ \ \text{for $a_{1}\neq a_{2}\neq...\neq a_{n}$}.
\end{equation}
If $b$ is different from all $a_{i}$, then $\gamma^{a_{1}...a_{n}}\gamma^{b}=\gamma^{a_{1}...a_{n}b}$. If $b=a_{i}$, we have
\begin{eqnarray}
&&\gamma^{a_{1}...a_{n}}\gamma^{b}=\gamma^{a_{1}}...\gamma^{a_{n}}\gamma^{b}\nonumber\\
&=&(-1)^{n+i}\gamma^{a_{1}}...\gamma^{a_{i-1}}\gamma^{a_{i+1}}...\gamma^{a_{n}}\gamma^{a_{i}}\gamma^{b}\nonumber\\
&=&(-1)^{n+i}\gamma^{a_{1}...a_{i-1}a_{i+1}...a_{n}}\eta^{a_{i}b},
\end{eqnarray}
and thus the first identity is proved. The second set of identities can be proved by using the first one twice.
\bigskip

\section{SUSY invariance of the action} \label{SUSYtransderive}

\begin{widetext}
We first derive the explicit form of $\Gamma^{a}(\epsilon)$ under SUSY
transformation. We denote 
\begin{eqnarray}
S_{0}&=&  -\frac{1}{12} e^{a}\wedge\overline{\psi} \gamma^{bcd}\wedge D\psi +\frac{\lambda}{8}e^{a}\wedge e^{b}\wedge\overline{\psi} \gamma^{cd}\wedge \psi -\frac{\lambda^{2}}{2} e^{a}\wedge e^{b}\wedge e^{c}\wedge e^{d}\equiv  S_{3/2}+S_{3}+S_{4}. \label{fourtermofS0}
\end{eqnarray}
Under the SUSY transformation Eq. (\ref{SUSYtrans}), we have 
\begin{eqnarray}
\delta  S_{3/2}
&=&-\frac{1}{12}\int \varepsilon_{abcd} [\frac{1}{2}e^{a}\wedge\overline{\psi} \gamma^{bcd} \gamma_{mn}\epsilon\wedge R^{mn}+2\lambda e^{a}\wedge\overline{\psi} \gamma^{bcd} \wedge D(e^{f}\gamma_{f}\epsilon)\nonumber\\
&&+T^{a}\wedge\overline{\psi} \gamma^{bcd}\wedge(D\epsilon+\lambda e^{f}\gamma_{f}\epsilon)-j^{a}\wedge\overline{\epsilon} \gamma^{bcd}\wedge D\psi+6 e^{a}\wedge j^{b}\wedge \delta\omega^{cd}]\nonumber\\
&=&\frac{1}{12}\int \varepsilon_{abcd} \left(3R^{\ d}_{m}\wedge e^{a}\wedge\overline{\psi} \gamma^{bcm}\epsilon + 3R^{cd}\wedge e^{a}\wedge\overline{\psi}\gamma^{b}\epsilon-DT^{a}\wedge\overline{\psi} \gamma^{bcd}\wedge\epsilon-T^{a}\wedge\overline{\psi}\overleftarrow{D} \gamma^{bcd}\wedge\epsilon +j^{a}\wedge\overline{\epsilon} \gamma^{bcd}\wedge D\psi  \right)\nonumber\\
&&+\frac{\lambda}{12}\int\varepsilon_{abcd}(T^{a}\wedge\overline{\psi} \gamma^{bcd}\gamma_{f}\epsilon \wedge e^{f}-2e^{a}\wedge\overline{\psi} \overleftarrow{D}\gamma^{bcd}\gamma_{f}\epsilon \wedge e^{f}-6\overline{\psi}\gamma^{ab}\epsilon\wedge j^{c}\wedge e^{d})\nonumber\\
&=&-\frac{1}{4}\int \varepsilon_{abcd}R^{ab}\wedge e^{c}\wedge\overline{\epsilon}\gamma^{d}\psi -\frac{1}{12}\int \varepsilon_{abcd}\left(T^{a}-j^{a}\right)\wedge\left(\overline{\epsilon} \gamma^{bcd}\wedge D\psi -\lambda\overline{\psi} \gamma^{bcd}\gamma_{f}\epsilon \wedge e^{f}\right)\nonumber\\
&&-\frac{\lambda}{2}\int\varepsilon_{abcd}\overline{\psi}\gamma^{ab}\epsilon\wedge j^{c}\wedge e^{d}+\frac{\lambda}{2}\int\varepsilon_{abcd}e^{a}\wedge e^{b}\wedge\overline{\epsilon}\gamma^{cd}D\psi.
\end{eqnarray}
In the first step, we use integral by parts for $\left(e^{a}\wedge\delta\overline{\psi} \gamma^{bcd}\wedge D\psi\right)$ and the identity $DD\epsilon=\frac{1}{4}R^{mn}\gamma_{mn}\epsilon$. For the product $\gamma^{bcd}\gamma_{mn}$ we have used Eq. (\ref{grassmannidentity2}).  For $\frac{1}{2}\varepsilon_{abcd}\overline{\epsilon}\gamma^{a}\psi\wedge\overline{\psi} \gamma^{bcd}\wedge D\psi$, we use Fierz rearrangement (the spacetime indices $\mu,\nu,...$ are total antisymmetric below):
\begin{eqnarray}
\frac{1}{2}\varepsilon_{abcd}\overline{\epsilon}\gamma^{a}\psi_{\mu}\overline{\psi}_{\nu} \gamma^{bcd}D_{\rho}\psi_{\sigma}
&=&-\frac{1}{8}\varepsilon_{abcd}[\overline{\psi}_{\nu}\gamma^{m}\psi_{\mu}\overline{\epsilon}\gamma^{a}\gamma_{m}\gamma^{bcd}D_{\rho}\psi_{\sigma}-\frac{1}{2}\overline{\psi}_{\nu}\gamma^{mn}\psi_{\mu}\overline{\epsilon}\gamma^{a}\gamma_{mn}\gamma^{bcd}D_{\rho}\psi_{\sigma}]\nonumber\\
&=&-\frac{1}{8}\frac{i}{6}[\overline{\psi}_{\nu}\gamma^{m}\psi_{\mu}\overline{\epsilon}\gamma^{a}\gamma_{m}\gamma_{a}\gamma_{5} D_{\rho}\psi_{\sigma}-\frac{1}{2}\overline{\psi}_{\nu}\gamma^{mn}\psi_{\mu}\overline{\epsilon}\gamma^{a}\gamma_{mn}\gamma_{a}\gamma_{5}D_{\rho}\psi_{\sigma}]\nonumber\\
&=&\frac{1}{4}\frac{i}{6}\overline{\psi}_{\nu}\gamma^{m}\psi_{\mu}\overline{\epsilon}\gamma_{m}\gamma_{5} D_{\rho}\psi_{\sigma}=-\frac{1}{4}\varepsilon_{abcd}\overline{\psi}_{\mu}\gamma^{a}\psi_{\nu}\overline{\epsilon}\gamma^{bcd}D_{\rho}\psi_{\sigma}. \label{jinS}
\end{eqnarray}
\end{widetext}
For the first line, due to the antisymmetry of the indices $\mu$ and $\sigma$, only these two terms survive.  The minus sign of the third line on the right hand side is because we change the order of $\gamma_{ab}$ comparing to Eq. (\ref{set}), and the additional coefficient $1/2$ is due to the repeated summation.  In the third step, we use integral by parts for $e^{a}\wedge\overline{\psi} \gamma^{bcd} \wedge D(e^{f}\gamma_{f}\epsilon)$. In the last step, we used that $DT^a=R^{a}_{\ f}\wedge e^{f}$ and
\begin{eqnarray}
&&\varepsilon_{abcd}R^{\ d}_{m}+\varepsilon_{abdm}R^{\ d}_{c}+\varepsilon_{adcm}R^{\ d}_{b}+\varepsilon_{dbcm}R^{\ d}_{a}=0\nonumber\\
& \Rightarrow&\ \ \ \varepsilon_{abcd}R^{\ d}_{m}\wedge e^{a}\gamma^{bcm}=\frac{1}{3} \varepsilon_{abcd} R^{a}_{\ f}\wedge e^{f} \gamma^{bcd}.
\end{eqnarray}
We also add a term $ -\frac{\lambda}{12}\int \varepsilon_{abcd} j^{a}\wedge \overline{\psi} \gamma^{bcd}\gamma_{f}\epsilon \wedge e^{f}$ in the last step since it is vanishing (easy to proved by using Fierz rearrangement, following the similar steps as Eq. (\ref{jinS})). And for the term $-\frac{\lambda}{6}\int\varepsilon_{abcd}e^{a}\wedge\overline{\psi} \overleftarrow{D}\gamma^{bcd}\gamma_{f}\epsilon \wedge e^{f}$, it can be written as: 
\begin{eqnarray}
&&-\frac{\lambda}{6}\int\varepsilon_{abcd}e^{a}\wedge\overline{\psi} \overleftarrow{D}\gamma^{bcd}\gamma_{f}\epsilon \wedge e^{f} \nonumber\\
&=&\frac{\lambda}{2}\int\varepsilon_{abcd}e^{a}\wedge e^{b}\wedge\overline{\epsilon}\gamma^{cd}D\psi,
\end{eqnarray}
where we use Eq. (\ref{grassmannidentity2}). The rank $4$ gamma matrix is vanishing due to $\varepsilon_{abcd}\gamma^{bcdf}\propto \delta^{f}_{a}$ and  $e^{a}\wedge e_{a}=0$. 

For the variation of $S_{3}$, we have
\begin{widetext}
\begin{eqnarray}
\delta  S_{3}&=& \frac{\lambda}{8}\int \varepsilon_{abcd}\overline{\epsilon}\gamma^{a}\psi\wedge e^{b}\wedge\overline{\psi} \gamma^{cd}\wedge \psi+ \frac{\lambda}{4}\int \varepsilon_{abcd}e^{a}\wedge e^{b}\wedge\overline{\psi} \gamma^{cd}\wedge (D\epsilon+\lambda e^{f}\gamma_{f}\epsilon)\nonumber\\
&=&\frac{\lambda}{8}\int \varepsilon_{abcd}\left( \overline{\epsilon}\gamma^{a}\psi\wedge e^{b}\wedge\overline{\psi} \gamma^{cd}\wedge \psi+ 4T^{a}\wedge e^{b}\wedge\overline{\psi} \gamma^{cd}\epsilon-2 e^{a}\wedge e^{b}\wedge\overline{\epsilon}\gamma^{cd}D\psi+4\lambda e^{a}\wedge e^{b}\wedge\overline{\psi}\gamma^{c}\epsilon\wedge e^{d}\right) \nonumber\\
&=&\frac{\lambda}{2} \int \varepsilon_{abcd}(T^{a}-j^{a})\wedge e^{b}\wedge\overline{\psi} \gamma^{cd}\epsilon -\frac{\lambda}{4}\int\varepsilon_{abcd}e^{a}\wedge e^{b}\wedge\overline{\epsilon}\gamma^{cd}D\psi +\frac{\lambda^{2}}{2}\int\varepsilon_{abcd}e^{a}\wedge e^{b}\wedge\overline{\psi}\gamma^{c}\epsilon\wedge e^{d},
\end{eqnarray}
\end{widetext}
where in the last line, we use the identity:
\begin{equation}
\frac{1}{4}\varepsilon_{abcd}\overline{\epsilon}\gamma^{a}\psi\wedge e^{b}\wedge\overline{\psi} \gamma^{cd}\wedge \psi=-\frac{1}{4}\varepsilon_{abcd}\overline{\psi}\gamma^{a}\psi\wedge e^{b}\wedge\overline{\psi} \gamma^{cd}\epsilon .
\end{equation}
This can be derived by using Fierz rearrangement similar as Eq. (\ref{jinS}).

For the variation of $S_{4}$, we have:
\begin{equation}
\delta  S_{4}=\lambda^{2}\int\varepsilon_{abcd}e^{a}\wedge e^{b}\wedge\overline{\psi}\gamma^{c}\epsilon\wedge e^{d}.
\end{equation}
Together we obtain the total variation of $S_0$:
\begin{widetext}
\begin{eqnarray}
\delta  S_{0}&=&\frac{1}{12}\int \varepsilon_{abcd}(T^{a}-j^{a})\wedge(-\overline{\epsilon}\gamma^{bcd}D\psi+\lambda\overline{\epsilon}\gamma^{bcdf}\psi\wedge e_{f}-3\lambda e^{b}\wedge\overline{\epsilon}\gamma^{cd}\psi)\nonumber\\
&&+\frac{\lambda}{4}\int \varepsilon_{abcd} e^{a}\wedge e^{b} \wedge \overline{\epsilon}\gamma^{cd} \left( D\psi-\lambda \gamma^{f}\psi\wedge e_{f}\right)- \frac{1}{4}\int \varepsilon_{abcd}e^{c}\wedge\overline{\epsilon}\gamma^{d}\psi \wedge \left( R^{ab}+\frac{\lambda}{2}\overline{\psi} \gamma^{ab}\wedge \psi+4\lambda^{2}e^{a}\wedge e^{b}\right).  \nonumber\\
 \label{deltaS0}
\end{eqnarray}
\end{widetext}
The first line gives the explicit form of $\Gamma_{a}(\epsilon)$ is: 
\begin{equation}
\Gamma_{a}(\epsilon)=\frac{1}{6}\varepsilon_{abcd}(-\overline{\epsilon}\gamma^{bcd}D\psi+\lambda\overline{\epsilon}\gamma^{bcdf}\psi\wedge e_{f}-3\lambda e^{b}\wedge\overline{\epsilon}\gamma^{cd}\psi).
\end{equation}

In order to cancel out this variation, we need to introduce three flat curvature constraint:
\begin{eqnarray}
&&\frac{1}{2}\int \widetilde{B}_{a}\wedge \left(T^{a}-j^{a}\right) ,\ \ \ \ \ \ 
\frac{1}{2}\int c\wedge \left( D\psi-\lambda \gamma^{f}\psi\wedge e_{f}\right) ,\nonumber\\
&&\frac{1}{2}\int B_{ab} \wedge \left( R^{ab}+\frac{\lambda}{2}\overline{\psi} \gamma^{ab}\wedge \psi+4\lambda^{2}e^{a}\wedge e^{b}\right).
\end{eqnarray}
At the first glance, we should choose the variation of $\widetilde{B}_{a}$ as:
\begin{equation}
\delta\widetilde{B}_{a}=-\Gamma_{a}(\epsilon)
\end{equation}
to cancel the corresponding term in $\delta S_{0}$. The total variation of $S_{\widetilde{B}}$ reads:
\begin{eqnarray}
\delta S_{\widetilde{B}}&=&\frac{1}{4 }\int \widetilde{B}_{a} \wedge \left(\overline{\epsilon}\gamma^{a}D\psi+\lambda\overline{\epsilon}\gamma^{a}\gamma^{b}e_{b}\wedge \psi \right) \nonumber\\
&&+\frac{1}{2}\int\delta \widetilde{B}_{a}\wedge \left(T^{a}-j^{a}\right) .  \label{deltaSBwide}
\end{eqnarray}
Thus we can choose the variation of $\overline{c}$ as $\delta\overline{c}= \frac{1}{2}\widetilde{B}_{a}\overline{\epsilon}\gamma^{a}-\frac{\lambda}{4} \varepsilon_{abcd} e^{a}\wedge e^{b} \wedge \overline{\epsilon}\gamma^{cd} $ to cancel out the corresponding term. Then total variation of $S_c$ reads:
\begin{widetext}
\begin{eqnarray}
\delta S_{c}&=& \frac{1}{2 }\int \delta \overline{c}\wedge \left( D\psi-\lambda \gamma^{a}\psi\wedge e_{a}\right) +\frac{1}{8}\int \delta\omega_{ab}\wedge \overline{c} \gamma^{ab}\wedge\psi +\frac{1}{8}\int \overline{c} \gamma^{ab}\epsilon\wedge R_{ab}\nonumber\\
&&+\frac{\lambda}{2}\int  \left(\overline{c}\gamma^{a}\epsilon\wedge T_{a}+\overline{c}\gamma^{a}\wedge D\epsilon\wedge e_{a}\right)-\frac{\lambda}{2}\int  \left(\overline{c}\gamma^{a}\wedge D\epsilon\wedge e_{a} +\lambda \overline{c}\gamma^{a}\gamma^{b}\epsilon\wedge e_{b}\wedge e_{a}-\frac{1}{2} \overline{c}\gamma^{a}\wedge\psi\wedge \overline{\psi}\gamma_{a}\epsilon \right) \nonumber\\
&=&\frac{1}{2 }\int \delta \overline{c}\wedge \left( D\psi-\lambda \gamma^{a}\psi\wedge e_{a}\right) +\frac{1}{8}\int \overline{c} \gamma^{ab}\epsilon\wedge (R_{ab}+4\lambda^2 e_{a}\wedge e_{b})+\frac{\lambda}{16}\int \overline{c} \gamma^{ab}\epsilon\wedge \overline{\psi}\gamma_{ab}\wedge\psi +\frac{\lambda}{2}\int \overline{c}\gamma^{a}\epsilon\wedge (T_{a}-j_{a})\nonumber\\
&=&\frac{1}{2 }\int \delta \overline{c}\wedge\left( D\psi+\lambda \gamma^{a}e_{a}\wedge \psi\right)+\frac{\lambda}{2}\int \overline{c}\gamma^{a}\epsilon\wedge (T_{a}-j_{a})+\frac{1}{2}\int \overline{c}\gamma_{ab}\epsilon\wedge \left(\frac{1}{4} R^{ab}+\frac{\lambda}{8}\overline{\psi} \gamma^{ab}\wedge \psi+\lambda^{2}e^{a}\wedge e^{b}\right), \label{dSc}\nonumber\\
\end{eqnarray}
\end{widetext}
where in the second equation we use Fierz rearrangement:
\begin{eqnarray}
(\overline{\psi}_{\mu}\gamma_{ab}\epsilon)( \overline{c}_{\nu\rho} \gamma^{ab}\psi_{\sigma})&=& \frac{1}{2}(\overline{\psi}_{\mu}\gamma^{ab}\psi_{\sigma})(\overline{c}_{\nu\rho} \gamma_{ab}\epsilon ). \nonumber\\
\end{eqnarray} 
In the derivation we have used the identities:
\begin{eqnarray}
&&\gamma^{a}\gamma_{m}\gamma_{a}=-2\gamma_{m}, \ \ \ \ \ \ \gamma^{a}\gamma_{mn}\gamma_{a}=0, \label{basicgamma} \nonumber\\ 
&&\gamma^{ab}\gamma^{m}\gamma_{ab}=0,\ \ \gamma^{ab}\gamma^{mn}\gamma_{ab}=4\gamma^{mn}. \label{twoid2}
\end{eqnarray}
Similarly, we also have
\begin{eqnarray}
(\overline{c}_{\mu\nu}\gamma^{a}\psi_{\rho})(\overline{\psi}_{\sigma}\gamma_{a}\epsilon )&=&-\frac{1}{2}(\overline{\psi}_{\rho}\gamma^{a}\psi_{\sigma})(\overline{c}_{\mu\nu} \gamma_{a}\epsilon )\nonumber\\
&=&-2j^{a}_{\ \rho\sigma}(\overline{c}_{\mu\nu} \gamma_{a}\epsilon ).
\end{eqnarray}
 We can see that the second term in the last line of Eq. (\ref{dSc}) can be absorbed by $\delta\widetilde{B}_{a}$ if we redefine the variation $\delta\widetilde{B}_{a}=-\Gamma_{a}(\epsilon)-\lambda\overline{c}\gamma^{a}\epsilon$. To cancel out the third term, we can choose the variation of $B_{ab}$ as $\delta B_{ab}=\frac{1}{4} \overline{c}\gamma_{ab}\epsilon + \frac{1}{4} \varepsilon_{abcd}e^{c}\wedge\overline{\epsilon}\gamma^{d}\psi$. The total variation of $S_{B}$ reads:
\begin{eqnarray}
\delta S_{B}&=& \frac{1}{2}\int \delta B_{ab} \wedge \left( R^{ab}+\frac{\lambda}{2}\overline{\psi} \gamma^{ab}\wedge \psi+4\lambda^{2}e^{a}\wedge e^{b}\right) \nonumber\\
&&-\frac{\lambda}{2}\int B_{ab} \wedge [\overline{\epsilon}\gamma^{ab} (D\psi+\lambda\gamma^{c}e_{c}\wedge\psi) ] .
\end{eqnarray}
The last line can also be absorbed by redefining $\delta \overline{c}=\frac{1}{2}\widetilde{B}_a\gamma^{a}\epsilon-\lambda B_{ab}\gamma^{ab}\epsilon+\frac{\lambda}{4}\varepsilon_{abcd}e^{a}\wedge e^{b}\gamma^{cd}\epsilon$. Finally the variation of auxiliary field can be determined as:
\begin{eqnarray}
&&\delta B^{ab}=-\frac{1}{4}\overline{c}\gamma^{ab}\epsilon+\frac{1}{2}\varepsilon^{abcd}e_{c}\wedge \overline{\epsilon}\gamma_{d}\psi,\nonumber\\
&&\delta c=\frac{1}{2}\widetilde{B}_a\gamma^{a}\epsilon-\lambda B_{ab}\gamma^{ab}\epsilon+\frac{\lambda}{4}\varepsilon_{abcd}e^{a}\wedge e^{b}\gamma^{cd}\epsilon\nonumber\\
&&\delta \widetilde{B}^{a}=-\Gamma^{a}(\epsilon)-\lambda \overline{c}\gamma^{a}\epsilon.
\end{eqnarray}
The SUSY transformation Eq. (\ref{SUSY1}) is exactly taking $\lambda=0$ in the variation of the above transformation. 

\section{ One-loop effective action} \label{piofpsi}

After the bosonic degrees of freedom acquire non-zero VEVs, we have effectively chosen a gauge condition when we drop their fluctuations, and no additional gauge fixing is required\footnote{Discarding the fluctuations of the bosonic fields is equivalent to choosing a gauge. This approach does not require the introduction of additional gauge-fixing terms or ghost fields. This can be proven by restoring the fluctuations and directly employing the Faddeev-Popov construction}. This can be seen from the fact that their propagators are no longer singular,
analogous to the Higgs mechanism where the gauge fields acquire mass after the scalar field obtains a non-zero VEV.
Without imposing further gauge condition, we can directly perform the path integral over the fermionic
degrees of freedom. Here we adopt the method from \cite{DSspecturm}, which decomposes the gauge field into physical modes (transverse and traceless) and gauge modes (which is not vanishing when gauge symmetry is breaking). For instance, for the gravitino field, we can decompose it as follows:
\begin{eqnarray}
&&\psi_{\mu}=\varphi_{\mu}+\frac{1}{4}\gamma_{\mu}\psi,\ \ \ \ \ \ \gamma^{\mu}\varphi_{\mu}=0,\nonumber\\
&&\varphi_{\mu}=\varphi_{\mu}^{\perp}+(\nabla_{\mu}-\frac{1}{4}\gamma_{\mu}\cancel{\nabla})\xi,\ \ \ \nabla^{\mu}\varphi_{\mu}^{\perp}=0. \label{psi}
\end{eqnarray}
The total covariant derivative is defined for spinor as:
\begin{equation}
\nabla_{\mu}\psi_{\nu}=\partial_{\mu}\psi_{\nu}-\Gamma^{\lambda}_{\mu\nu}\psi_{\lambda}+\frac{1}{4}\gamma^{ab}\omega_{ab\mu}\psi_{\nu}.
\end{equation}
The constrained Laplacian $\triangle_{s}(X)$ can be defined as:
\begin{eqnarray}
&& \triangle_{1/2}(X)\psi=(-\cancel{\nabla}^2+X)\psi=(-\nabla^2+\Lambda+X)\psi,\nonumber\\
&& \triangle_{3/2}(X)\varphi_{\mu}^{\perp}=(-\cancel{\nabla}^2+X)\varphi_{\mu}^{\perp}=(-\nabla^2+ 	\frac{4}{3}\Lambda+X)\varphi_{\mu}^{\perp}. \label{constraint_operator}\nonumber\\
\end{eqnarray}
The spectrum of these  constrained operators in the AdS space are known\cite{ADSspecturm}. Substitute Eq. (\ref{tree_and_oneloop}) into the first term of $S_f$, we obtain
\begin{eqnarray}
&& S_{\psi_{\mu}}=-\frac{1}{2l_p}\int \overline{e}\ [\overline{\varphi}_{\mu}^{\perp}(\cancel{\nabla}+m_\psi)\varphi_{\mu}^{\perp}-\frac{3}{8}\overline{\psi}(\cancel{\nabla}-2m_\psi)\psi] \nonumber\\
&&-\frac{3}{16l_p}\int \overline{e}\ [\overline{\xi}(\cancel{\nabla}+2m_\psi)\triangle_{1/2}(-\frac{4}{3}\Lambda)\xi-2\overline{\xi}\triangle_{1/2}(-\frac{4}{3}\Lambda)\psi]. \nonumber
\end{eqnarray}
Such transformation will induce a Jacobian:
\begin{equation}
[D\psi]=J_{\psi}[D\varphi_{\mu}^{\perp}][D\xi][\nabla\psi],
\end{equation}
which can be calculated through
\begin{eqnarray}
1&=&\int [D\psi]\exp[-\frac{1}{2}\int \overline{e}\ \overline{\psi}_{\mu}\psi_{\mu}]\nonumber\\
&=& \int J_{\psi}[D\varphi_{\mu}^{\perp}][D\xi][D\psi]\exp[-\frac{1}{2}\int \overline{e}\ (\overline{\varphi}_{\mu}^{\perp}\varphi_{\mu}^{\perp}\nonumber\\
&&+\frac{1}{4}\overline{\psi}\psi-\overline{\xi}(\nabla^{\mu}+\frac{1}{4}\cancel{\nabla}\gamma^{\mu})(\nabla_{\mu}-\frac{1}{4}\gamma_{\mu}\cancel{\nabla})\xi)]\nonumber\\
&=& J_{\psi}[\det\triangle_{1/2}(-\frac{4}{3}\Lambda)]^{1/2} ,\label{jaco}
\end{eqnarray}
which yields:
\begin{equation}
J_{\psi}=[\det\triangle_{1/2}(-\frac{4}{3}\Lambda)]^{-1/2}.\label{Jacobian_of_psi}
\end{equation}
The above action can be diagonalized through transformation
\begin{eqnarray}
    \psi\rightarrow \psi+\cancel{\nabla}\xi+\frac{2(2\Lambda+3m_\psi\cancel{\nabla})}{3(\cancel{\nabla}-2m_\psi)}\xi,
\end{eqnarray}
to
\begin{eqnarray}
    S_{\psi_{\mu}}&=&-\frac{1}{2l_p}\int \overline{e}\ [\overline{\varphi}_{\mu}^{\perp}(\cancel{\nabla}+m_\psi)\varphi_{\mu}^{\perp}-\frac{3}{8}\overline{\psi}(\cancel{\nabla}-2m_\psi)\psi]   \nonumber\\
&&-\frac{3m_\psi^2+\Lambda}{4 l_p}\int e\ \overline{\xi}\frac{\cancel{\nabla}^2+4\Lambda/3}{\cancel{\nabla}-2m_\psi}\xi.
\end{eqnarray}
This shift of $\psi$ does not change the Jacobian (\ref{Jacobian_of_psi}). And integral over $\psi_\mu$ give rise to:
\begin{eqnarray}
Z_\psi=\int [D\psi]e^{iS_{\psi_\mu}}&=&\int J_{\psi}[D\varphi_{\mu}^{\perp}][D\xi][\nabla\psi]e^{iS_{\psi_\mu}}\nonumber\\
&=&[\det\triangle_{3/2}(m_\psi^2)]^{1/4}, \label{oneloop_of_psi}
\end{eqnarray} 
where we used the identity:
\begin{eqnarray}
    \det(\cancel{\nabla}+m_\psi)^{1/2}\det(\cancel{\nabla}+m_\psi)^{1/2}=\det(\cancel{\nabla}^2-m_{\psi}^2)^{1/2}, \nonumber
\end{eqnarray}
which can be proved by multiplying both sides of the operator by $\gamma_5$ and then moving the leftmost $\gamma_5$ to the rightmost position.

After integrating over $\psi_\mu$, the 2-form fields $c_{\mu\nu}$ acquire quadratic action:
\begin{eqnarray}
    S_{c_{\mu\nu}}=-\frac{i}{2}\LL (\frac{i}{4}\int \varepsilon^{\mu\nu\rho\sigma}\overline{c}_{\mu\nu}(\nabla_{\rho}+m_{c}\gamma_{\rho})\psi_{\sigma})^2 \RR_\psi, \label{Sc_average_form}
\end{eqnarray}
where the subscript $\psi$ stands for the average with respect to $\psi$, i.e., $\LL O \RR_\psi=\int D\psi O \exp(i S_\psi)/\int D\psi\exp(i S_\psi)$. Performing the path integral for spin-1/2 fields is relatively straightforward because the inverse of the Dirac operator is well-known. However, the propagator for spin-3/2 fields is more complicated due to the constraints on the field $\varphi_{\mu}^{\perp}$.   we suppose that the propagator $\langle \varphi^{\perp}_{\mu}(x)\overline{\varphi}^{\perp}_{\nu}(y) \rangle$ can be written as 
\begin{equation}
\langle \varphi^{\perp}_{\mu}(x)\overline{\varphi}^{\perp}_{\nu}(y) \rangle=\frac{\delta(x-y)}{e}\Pi_{\mu\nu}(\nabla,g,\gamma)
\end{equation}
with $\Pi_{\mu\nu}(\nabla,g,\gamma)$ an operator contains covariant derivative and satisfies
\begin{eqnarray}
\nabla^{\mu}\Pi_{\mu\nu}\psi^{\nu}=\overline{\psi}^{\mu}\overleftarrow{\Pi}_{\mu\nu}\overleftarrow{\nabla}^{\nu}=\gamma^{\mu}\Pi_{\mu\nu}\psi^{\nu}=\overline{\psi}^{\mu}\Pi_{\mu\nu}\gamma^{\nu}=0 \nonumber
\end{eqnarray} 
for arbitrary spinor-vector $\psi_{\mu}$. The arrow over the operator denotes that the derivative acting to the left. After tedious calculation, we find that 
\begin{eqnarray}
\Pi_1&=& \nabla^2 g_{\mu\nu}-\frac{1}{2} (\nabla_{\mu}\nabla_{\nu}+\nabla_{\nu}\nabla_{\mu})-\frac{1}{2}  \nabla^2 \gamma_{\mu\nu}\nonumber\\
&&-\frac{1}{2}  (\nabla_{[\mu} \gamma_{\nu]\rho}\nabla^{\rho}+\nabla^{\rho}\nabla_{[\mu} \gamma_{\nu]\rho})-\frac{\Lambda}{4} \Lambda g_{\mu\nu}+\frac{\Lambda}{12}\Lambda \gamma_{\mu\nu} ,\nonumber\\
\Pi_2 &=& g_{\mu\nu} (\nabla^2\slashed{\nabla}+\slashed{\nabla}\nabla^2+\frac{5\Lambda}{6}\slashed{\nabla})- (\nabla_{\mu}\slashed{\nabla}\nabla_\nu+\nabla_{\nu}\slashed{\nabla}\nabla_\mu) \nonumber\\
&&-\frac{1}{2} (\nabla^2\nabla^{\rho}+\nabla^{\rho}\nabla^2+\frac{4\Lambda}{3}\nabla^{\rho})\gamma_{\mu\nu\rho} +\frac{\Lambda}{3}\nabla_{(\mu}\gamma_{\nu)} \nonumber
\end{eqnarray}
satisfy these constraints. The overall coefficient can be determined by the propagator of $\varphi_{\mu}^{\perp}$ in the flat spacetime:
\begin{eqnarray}
&&\Pi_{\mu\nu}^{\text{flat}}=iP_{\mu}^{\ \rho}(-i\slashed{p}+m_\psi)^{-1}_{\rho\sigma}P^{\sigma}_{\ \nu} \nonumber\\
&=&\frac{2i}{3
p^2(p^2+m_{\psi}^{2})}[m_{\psi}(p^{2}\eta_{\mu\nu}-p_{\mu}p_{\nu}-\frac{p^{2}}{2}\gamma_{\mu\nu}-p_{[\mu}\gamma_{\nu]\rho}p^{\rho})\nonumber\\
&&-i(p^{2}\eta_{\mu\nu}\cancel{p}-p_{\mu}p_{\nu}\cancel{p}-\frac{p^{2}}{2}\gamma_{\mu\nu\rho}p^{\rho})], \label{flat_space_pro} \nonumber
\end{eqnarray}
where $P_{\mu}^{\ \rho}$ is the projection operator
\begin{eqnarray}
P_{\mu\nu}=\frac{1}{3}(2\eta_{\mu\nu}-\frac{2p_\mu p_\nu}{p^2}  -\gamma_{\mu\nu}-\frac{2p_{[\mu}\gamma_{\nu]\rho}p^{\rho}}{p^2}),
\end{eqnarray}
which satisfies 
\begin{equation}
P^{\mu}_{\ \rho}P^{\rho}_{\ \nu}=P^{\mu}_{\ \nu},\ \ \  \gamma_{\mu}P^{\mu}_{\ \nu}=p_{\mu}P^{\mu}_{\ \nu}=P^{\mu}_{\ \nu}\gamma^\nu=P^{\mu}_{\ \nu}p^\nu=0. \nonumber
\end{equation}
Thus the propagator of $\varphi_{\mu}^{\perp}$ takes the form
\begin{equation}
\langle \varphi^{\perp}_{\mu}(x)\overline{\varphi}^{\perp}_{\nu}(y) \rangle=\frac{\delta(x-y)}{e}\frac{2i}{3 \slashed{\nabla}^2(m_\psi^2-\slashed{\nabla}^2)}(m_\psi \Pi_1+\frac{1}{2}\Pi_2).
\end{equation}
 For the new ingredient $c_{\mu\nu}$, we decompose them into irreducible representation as:
\begin{eqnarray}
c_{\mu\nu}=2\nabla_{[\mu}k_{\nu]}^{\perp}+2\gamma_{[\mu}b_{\nu]}^{\perp}+2\gamma_{[\mu}\nabla_{\nu]}\chi+\gamma_{\mu\nu}c, \label{decompose_of_c}
\end{eqnarray}
with
\begin{eqnarray}
 D_{\mu}k_{\mu}^{\perp}=\gamma_{\mu}k_{\mu\nu}^{\perp}=0,\ \ D_{\mu}b_{\mu}^{\perp}=\gamma_{\mu}b_{\mu}^{\perp}=0.
\end{eqnarray}
Such decomposition induce the Jacobian
\begin{equation}
J_{c}=[\det\triangle_{3/2}(-\frac{1}{3}\Lambda)\det\triangle_{1/2}(-\frac{4}{3}\Lambda)]^{-1/2}. \label{Jacobian_of_c}
\end{equation}
The constraint propagator of  $\varphi_{\mu}^{\perp}$ does not  contribute to the induced quadratic action of spin 1/2 fields, i.e.,
\begin{eqnarray}
\overline{c}\langle \varphi^{\perp}\overline{\varphi}^{\perp} \rangle c=\overline{\chi}\langle \varphi^{\perp}\overline{\varphi}^{\perp} \rangle \chi=0.
\end{eqnarray}
The spin $1/2$ propagator also does not contribute to the induced quadratic action of the spin $3/2$ fields,
\begin{eqnarray}
\overline{b}^{\perp}\langle \xi\overline{\xi} \rangle b^{\perp}=\overline{k}^{\perp}\langle \xi\overline{\xi} \rangle k^{\perp}=\overline{b}^{\perp}\langle \psi\overline{\psi} \rangle b^{\perp}=0.
\end{eqnarray}
Note that $\psi_\mu$ in Eq. (\ref{Sc_average_form}) has been decomposed into 
\begin{eqnarray}
\psi_{\mu}=\varphi_{\mu}^{\perp}+(\nabla_{\mu}+\frac{1}{6}\gamma_{\mu}\frac{2\Lambda-3m_\psi\cancel{\nabla}}{\cancel{\nabla}+2m_\psi})\xi+\frac{1}{4}\gamma_{\mu}\psi.
\end{eqnarray}
We could simplify the calculation by utilizing the fact that the constraint fields are eigenstate of Dirac operator, i.e.,
\begin{eqnarray}
    \slashed{\nabla}\psi_n=\lambda_n\psi_n,\ \ \  \slashed{\nabla}\varphi^{\perp}_{n\mu}=\lambda'_n\varphi^{\perp}_{n\mu}.
\end{eqnarray}
The quadratic action of $c_{\mu\nu}$ after integrating out $\varphi_{\mu}^{\perp}$, $\xi$ and $\psi$ reads:
\begin{eqnarray}
S_{c_{\mu\nu}}=\int \overline{e} &(\overline{b}_\mu^{\perp}O_1 b^{\perp\mu}+\overline{b}_\mu^{\perp}O_2 k^{\perp\mu}+\overline{k}_\mu^{\perp}O_3 k^{\perp\mu} \nonumber\\
&+\overline{c} G_1 c+\overline{c} G_2 \chi+\overline{\chi} G_3 \chi). 
\end{eqnarray}
The explicit forms of these operators are rather complicated, so we will not provide them individually. Ultimately, the result of the path integral only depends on the discriminant of the quadratic form, which has a relatively simple form, as follows:
\begin{eqnarray}
  && O_2.O_2- 4O_1.O_3=\nonumber\\
  &&\ \ \ \  \frac{64\Lambda(\Lambda+12m_c^2)^2(\slashed{\nabla}^2+\frac{13}{6}\Lambda)(\slashed{\nabla}^2+\frac{1}{3}\Lambda) }{81\slashed{\nabla}^3(m_\psi^2-\slashed{\nabla}^2)(\slashed{\nabla}-m_\psi)} , \label{c01} \\
  && G_2.G_2  -4G_1.G_3=\frac{16(\Lambda+12m_c^2)^2(-\slashed{\nabla}^2-\frac{4}{3}\Lambda)}{\Lambda+3m_\psi^2}.\nonumber\\ \label{co2}
\end{eqnarray}
Integrating out these component, we obtain the effective one-loop action contribute by $c_{\mu\nu}$:
\begin{eqnarray}
Z_c&=&\int [Dc]e^{iS_{c_{\mu\nu}}}  \nonumber\\
&=&J_c\times(O_2.O_2- 4O_1.O_3)^{1/2}  ( G_2.G_2-4G_1.G_3)^{1/2}  \nonumber\\
&=&\frac{[\det\triangle_{3/2}(-\frac{13\Lambda}{6})]^{1/2}}{[\det\triangle_{3/2}(0)\det\triangle_{3/2}(m_\psi^2)]^{3/4}},  \label{oneloop_of_c}
\end{eqnarray}
where we have dropped some constant factors.

\section{Zeta functions and self-consistent equations in the AdS spacetime}\label{zeta_and_self}

When computing the one-loop effective action, we perform a Wick rotation to Euclidean space to ensure convergence of the integrals. It is important to note that in doing the Wick rotation, we do not change the underlying spacetime manifold itself, but rather transform the 00 component of the emergent background metric $\overline{g} _{00}\rightarrow -\overline{g} _{00}$, or equivalently, the vierbein $\overline{e} ^{0}_{\mu}\rightarrow -i\overline{e} ^{0}_{\mu}$.\footnote{Here the minus sign is chosen such the scalar field $-\int d^4x (\partial_\mu \phi)^2$ has right sign.}   Zeta function is defined through the eigenvalues of operator $\triangle_{s}$:
\begin{equation}
    \triangle_{s}(X)\phi_n=\lambda_n\phi_n,\ \ \ \zeta^{(s)}(p)=\sum_n \lambda_n^{-p}.
\end{equation}
 Since Ads space is non-compact, the eigenvalues of the
operators are continuous, summation will covert to integration. Here we present the result of zeta functions in\cite{ADSspecturm}:
\begin{eqnarray}
    \zeta^{(s)}(0,b)&=&\frac{V(H_4)(2s+1)}{32\pi^2 a^4}[b^2-(s+\frac{1}{2})^2(2b-\frac{1}{3})+\frac{1}{30}],   \nonumber\\ 
    \zeta^{(s)'}(0,b)&=&\frac{V(H_4)(2s+1)}{32\pi^2 a^4}[b^2-\frac{4}{3}b^{3/2}-\frac{1}{3}b+(2s+1)^2\sqrt{b}   \nonumber\\ 
    &&-8c+8\int_0^{\sqrt{b}}[(s+\frac{1}{2})^2-\lambda^2]\psi(\lambda)\lambda d\lambda],
\end{eqnarray}
for $s=\frac{1}{2},\frac{3}{2},...$. The digamma function is defined as  $\psi(\lambda)=\Gamma'(\lambda)/\Gamma(\lambda)$ and the constant is
\begin{eqnarray}
    c=\int_0^\infty d\lambda \frac{\lambda^3+(s+\frac{1}{2})^2\lambda}{e^{2\pi\lambda}-1}\ln(\lambda^2).
\end{eqnarray}
The $\psi(x)$ function has following asymptotic behaviour:
        \begin{eqnarray}
\lim_{\substack{ |x|\rightarrow \infty, \\
\text{arg } x>-\pi+i\epsilon}}\psi(x)=\ln x+O(\frac{1}{x}),\ \ \lim_{|x|\rightarrow 0}\psi(x)=-\frac{1}{x}. \label{asymptotic_behaviour} \nonumber\\
\end{eqnarray}
We define a dimensionless parameter 
\begin{eqnarray}
    y=\lambda^2a^2/(3l_p^2),
\end{eqnarray}
which represents the ratio between the bare cosmological constant and the effective cosmological constant. We can prove that the original self-consistent Eq.s (\ref{self-consistent}) are equivalent to 
\begin{eqnarray}
    \frac{\delta S_{\text{eff}}}{\delta l_p}=\frac{\delta S_{\text{eff}}}{\delta B}=\frac{\delta S_{\text{eff}}}{\delta y}=0.  \label{self_consistent_y}
\end{eqnarray}
This can be proved by noticing that
\begin{equation}
    \frac{\delta}{\delta \overline{g}_{\mu\nu}}\left( \frac{V(H_4)}{a^4}\right)=0
\end{equation}
and using the chain rule. In order to solve the self-consistent Eq.s (\ref{self_consistent_y}), we consider different cases based on the range of values of $y$:
\begin{enumerate}
    \item For $y\rightarrow \infty$ or $|\Lambda|\ll \lambda^2/l_p^2$.  In this case $\zeta^{(s)}(0,b),\zeta^{(s)'}(0,b)$ in $\triangle_{3/2}(0)$ and $\triangle_{3/2}(-13\Lambda/6)$  are of order $O(1)$. However, there are some subtle problems in $\triangle_{3/2}(m_\psi^2)$ since $a^2m_\psi^2\rightarrow\infty$. Using asymptotic behavior Eq. (\ref{asymptotic_behaviour}), we find that the variation  of the last term in $\zeta^{(s)'}(0,b)$ with respect to $b$ cancels exactly with the variation  of $\zeta^{(s)}(0)\ln a^2$. Thus, the leading order term of $S^E_{\text{eff}}$ is 
  \begin{eqnarray}
   S^{(0)}&=&-\frac{V(H_4)}{a^4} \{\frac{108(2B-1)y^2}{\lambda^2}  \nonumber\\
  &&   +\frac{1}{16\pi^2}[(B+1)^4y^2\ln\frac{3l_p^2\mu^2y^2}{\lambda^2} \nonumber\\
  &&+ (B+1)^4y^2+I((B+1)^2y)
   ]\},
  \end{eqnarray}
where 
\begin{equation}
    I(x)=8\int_0^{\sqrt{x}}(4-\lambda^2)\psi(\lambda)\lambda d\lambda
\end{equation}
and $I'(x)=-2x\ln x+O(x^{-1})$. Combining $\delta S^{(0)}/\delta B=\delta S^{(0)}/\delta B=0$, we have
\begin{equation}
    B=2+O(y^{-1}).
\end{equation}
However, $\delta S^{(0)}/\delta l_p=0$ gives rise to
\begin{equation}
     B=-1+O(y^{-1}).
\end{equation}
Thus this case is invalid.

\item  For $y\rightarrow 0$ or $|\Lambda|\gg \lambda^2/l_p^2$. In this case the leading order term of the effective action can be written as
\begin{eqnarray}
   S^{(0)}&=&\frac{V(H_4)}{a^4} \{18By+c'\nonumber\\
  &&  + \frac{1}{16\pi^2}[I((B+1)^2y)-\frac{59}{30}\ln\frac{3l_p^2\mu^2y^2}{\lambda^2} 
   ]\},\nonumber\\
\end{eqnarray}
where we denote some unimportant constant by $c'$. In this case $I'(x)=16 x^{-1}+O(x)$.
The self-consistent equation $\delta S^E_{\text{eff}}/\delta l_p=0$ can't be satisfied in this case.

\item For $y\sim 1$. In this case $\zeta^{(s)}(0,b),\zeta^{(s)'}(0,b)$ in $\triangle_{3/2}(0)$, $\triangle_{3/2}(-13\Lambda/6)$ and $\triangle_{3/2}(m_\psi^2)$  are of order $O(1)$. However, the first term in Eq. (\ref{effect_action}) (tree-level contribution) can be rewritten as:
\begin{eqnarray}
    \frac{18 V(H_4)}{ a^4\lambda^2} 
    [ By-6(2B-1)y^2]
\end{eqnarray}
which is of order $\lambda^{-2}$. Initially, we assume that 
\begin{eqnarray}
    B=1+O(\lambda^2),\ \ \ y=\frac{1}{12}+O(\lambda^2).
\end{eqnarray}
This on-shell solution with very small quantum correction makes the contribution of the tree level term to self-consistent equations $\delta S_{\text{eff}}/\delta B$ and $\delta S_{\text{eff}}/\delta y$ of order $O(1)$, which is compatible with the one-loop correction. However, we find that this solution does not satisfy $\delta S_{\text{eff}}/\delta l_p=0$.  Therefore, to satisfy all  self-consistent equations, the tree-level contribution and quantum corrections are in the same order, i.e. $\ln(a^2\mu^2)\sim\ln(l_p^2\mu^2/\lambda^2)\sim \lambda^{-2}$, which implies that $l_p\sim \exp(g/\lambda^2)$ ($g>0$) is a very large scale. Under such a condition, we can neglect the contribution from $\zeta^{(s)'}(0)$. The effective action can be simplified as:
\begin{eqnarray}
   S^E_{\text{eff}}&=&\frac{3V(H_4)}{8\pi^2a^4}\{  \frac{48\pi^2 }{\lambda^2}[B y-6(2B-1)y^2]  \nonumber\\ 
    &&+\frac{1}{2}\ln\frac{3l_p^2\mu^2y^2}{\lambda^2}[ 8 (1 + B)^2 y - 3 (1 + B)^4 y^2-\frac{59}{15} ]\}.  \nonumber\\ 
\end{eqnarray}

\end{enumerate}
The corresponding self-consistent Eq. (\ref{self_consistent_y}) reads:
\begin{eqnarray}
  \frac{\delta S_{\text{eff}}}{\delta l_p}&=&\frac{3V(H_4)}{8\pi^2a^4l_p}[ 8 (1 + B)^2 y - 3 (1 + B)^4 y^2-\frac{59}{15} ]=0,\nonumber\\
  \frac{\delta S_{\text{eff}}}{\delta y}&=&\frac{3V(H_4)}{8\pi^2a^4}\{  \frac{48\pi^2 }{\lambda^2}[B -12(2B-1)y]  \nonumber\\ 
    &&+\frac{1}{2}\ln\frac{3l_p^2\mu^2y^2}{\lambda^2}[ 8 (1 + B)^2  - 6 (1 + B)^4 y ]\}=0,\nonumber\\
  \frac{\delta S_{\text{eff}}}{\delta B}  &=&\frac{3V(H_4)y}{8\pi^2a^4}\{  \frac{48\pi^2 }{\lambda^2}(1-12y)  \nonumber\\ 
    &&+2\ln\frac{3l_p^2\mu^2y^2}{\lambda^2}[ 4(1 + B) - 3 (1 + B)^3 y ]\}=0,  \nonumber\\ 
\end{eqnarray}
where in the second equation we drop the term coming from the derivative of the logarithm term with respect to $y$, since it is of order $O(1)$. For the case when $y\sim 1$, we have calculated the following physical solution:
\begin{eqnarray}
  &&  B=7.52,\ \ \ y=\frac{1}{36},  \nonumber\\
  &&  z=\frac{\lambda^2}{48\pi^2}\ln\frac{3l_p^2\mu^2y^2}{\lambda^2}=0.0191,  \label{non_trivial_solution}
\end{eqnarray}
Here, the physical solution means that $B>0$ and $z>0$. 
This is because  Eq. (\ref{S_B_final}) shows that $B$ is effectively the overall coefficient of the Einstein term, and we require it to be positive in order to maintain consistency with the conventional Einstein–Hilbert action. At the same time, $z>0$ implies that the VEV of $\overline{e}^a_\mu$ vanishes as $\lambda$ approaches 0. In this case, classical spacetime breaks down, which is consistent with the conclusion we obtained previously from our undeformed theory.

\section{Absence of saddle point solutions in the flat spacetime and dS spacetime}\label{flat_and_ads}

In this section, we shall show that flat spacetime and dS spacetime do not possess saddle points given by self-consistent Eq.s (\ref{self-consistent}).  Notice that the decomposition of $\psi_\mu$ and $c_{\mu\nu}$ in Eq. (\ref{psi}),(\ref{decompose_of_c}), as well as the Gaussian integrals, do not depend on the specific spacetime (since their irreducible representations always include transverse modes, longitudinal modes, and the trace part). Therefore, the expressions listed in Appendix \label{piofpsi} can still be applied here, provided that we account for the change in the effective cosmological constant.

For flat spacetime, the cosmological constant is vanishing. Notice that there is a factor $\Lambda$ in Eq. (\ref{c01}), which means the discriminant  is vanishing in the flat spacetime. In the flat spacetime, we calculated that 
\begin{eqnarray}
    &&O_1=-\frac{4(\slashed{\partial}-2m_c)^2}{\slashed{\partial}-m_\psi},\ \ O_2=\frac{8m_c\slashed{\partial}(\slashed{\partial}-2m_c)}{\slashed{\partial}-m_\psi},\nonumber\\
    &&\ \ \ \ \ \ \ \ \ \ \ \ \ \ \ \ \ \  O_3=-\frac{4\slashed{\partial}^2m_c^2}{\slashed{\partial}-m_\psi}.
\end{eqnarray}
The bilinear term of $3/2$ fields can be transformed to
\begin{eqnarray}
   && \overline{b}_\mu^{\perp}O_1 b^{\perp\mu}+\overline{b}_\mu^{\perp}O_2 k^{\perp\mu}+\overline{k}_\mu^{\perp}O_3 k^{\perp\mu} = \nonumber\\
  &&   -(\overline{b}-\frac{2\slashed{\partial}}{(\slashed{\partial}-2m_c)}\overline{k})_\mu^{\perp}\frac{4(\slashed{\partial}-2m_c)^2}{\slashed{\partial}-m_\psi}(b-\frac{2\slashed{\partial}}{(\slashed{\partial}-2m_c)} k)^{\perp\mu}. \nonumber
\end{eqnarray}
Thus the effective one-loop action contribute by $c_{\mu\nu}$
\begin{eqnarray}
Z_c&=&\int [Dc]e^{iS_{c_{\mu\nu}}}  \nonumber\\
&=&J_c \times (O_1)^{1/2} ( G_2.G_2-4G_1.G_3)^{1/2}  \nonumber\\
&=&\frac{[\det\triangle_{3/2}(0)]^{1/2}[\det\triangle_{3/2}(4m_c^2)]^{1/2}}{[\det\triangle_{3/2}(m_\psi^2)]^{1/4}}.  
\end{eqnarray}
Combining with the contribution from $\psi_\mu$ 
\begin{eqnarray}
    Z_\psi=[\det\triangle_{3/2}(m_\psi^2)]^{1/4},
\end{eqnarray}
we can see that the quantum correction does not depend on $B$. The total effective action reads:
\begin{eqnarray}
    S_{\text{eff}}&=&-\frac{2(2B-1)V\lambda^2}{l_p^4}   \nonumber\\ 
    &&-\frac{1}{2}\ln[\det\triangle_{3/2}(0)\det\triangle_{3/2}(4m_c^2)]. \label{effect_action_flat}
\end{eqnarray}
The self-consistent equation $\delta S_{\text{eff}}/\delta B=0$ requires that $l_p\rightarrow \infty$ which contradicts our assumption that $e^a_\mu$ has non-zero VEV.

For a dS spacetime, we can use the expression in Eq. (\ref{oneloop_of_c}), since $\Lambda$ is nonzero. The total effective action reads:
\begin{eqnarray}
    S_{\text{eff}}&=&-\frac{V(S_4) }{l_p^2}[\frac{2(2B-1)\lambda^2}{l_p^2}-\frac{B}{a^2}]   \nonumber\\ 
    &&-\frac{1}{2}\ln\frac{\det\triangle_{3/2}(-\frac{13}{6}\Lambda)}{\det\triangle_{3/2}(m_\psi^2)[\det\triangle_{3/2}(0)]^{3/2}}, \label{effect_action_ds}
\end{eqnarray}
where $V(S_4)$ is the volume of the 4 dimension sphere. Since the spacetime is compact, the spectrum of the operators $\triangle_s(X)$ is discrete. The explicit form of zeta functions in the dS space can be found in\cite{DSspecturm}
\begin{widetext}
    \begin{eqnarray}
\frac{3}{2s+1}\zeta_s(0,b_s)&=&\frac{1}{4}b_s(b_s-2a_s)+\frac{1}{24}a_s(3k_s^2+6k_s+2)-\frac{1}{64}k_s^2(k_s+2)^2+\frac{1}{120}, \nonumber\\    
\frac{3}{2s+1}\zeta'_s(0,b_s)&=&\frac{1}{4}b_s^2-\frac{1}{12}b_s-\frac{1}{8}b_sk_s(k_s+2)-\frac{1}{2}\int_0^{b_{s}} \ dz(z-a_s)[\psi(s+\frac{3}{2}+\sqrt{z})+\psi(s+\frac{3}{2}-\sqrt{z})]+c, \label{zeta_zeta'_ds}
\end{eqnarray}
\end{widetext}
where $a_s=(s+\frac{1}{2})^2,\ k_s=2s+1,\ b_s=a^2X$. In order to solve the self-consistent equations, we still consider the cases based on the range of $y=\lambda^2a^2/(3l_p^2)$. For the cases $y\rightarrow\infty$ and $y\rightarrow 0$, the discussion is identical to that in Appendix \ref{zeta_and_self}, where saddle points do not exist. As for $y\sim 1$, a solution is still possible only if $\ln (l_p^2\mu^2)\sim \lambda^{-2}$. In this case,  the effective action can be simplified as:
\begin{eqnarray}
    S^{E}_{\text{eff}}&=&-\frac{48\pi^2}{\lambda^2}[By-6(2B-1)y^2] \nonumber\\
    &&-\frac{1}{2}\ln\frac{3l_p^2\mu^2y^2}{\lambda^2}[ 8 (1 + B)^2 y - 3 (1 + B)^4 y^2-\frac{59}{15} ], \nonumber\\ 
\end{eqnarray}
where we substitute $V(S_4)=-8\pi^2a^2/3$ and remove the additional zero modes from the zeta functions of $\psi_\mu,\ c_{\mu\nu}$\cite{DSspecturm}:
\begin{eqnarray}
    &&\zeta(\psi_\mu)\rightarrow\zeta(\psi_\mu)-4, \nonumber\\
    &&\zeta(c_{\mu\nu})\rightarrow\zeta(c_{\mu\nu})+8 .
\end{eqnarray}
As discussed in the previous section, the self-consistent Eq.s (\ref{self-consistent}) and Eq.s (\ref{self_consistent_y}) are equivalent. However, in the dS spacetime, we only find the following solution:
\begin{eqnarray}
  &&  B=7.52,\ \ \ y=\frac{1}{36},  \nonumber\\
  &&  z=\frac{\lambda^2}{48\pi^2}\ln\frac{3l_p^2\mu^2y^2}{\lambda^2}=0.0191,  \label{non_trivial_solution}
\end{eqnarray}
This implies that the spacetime radius $a^2$ is positive, contradicting the dS spacetime assumption; hence no saddle-point solution exists.

\bibliography{cite.bib}

\end{document}